# Bayesian non-parametric lumping and splitting of nodes in Network Meta-Analysis under heterogeneity

Tim Disher; Chris Cameron, Brian Hutton

2025-06-27

## 1. Abstract

Network meta-analysis (NMA) synthesizes evidence for multiple treatments, but decisions on node formation can have important statistical implications including bias or inflated uncertainty. Existing data-driven methods often lack flexibility or fail to fully account for node uncertainty and adjust for between-trial heterogeneity simultaneously. We introduce a Bayesian non-parametric framework using a Dirichlet process prior with a regularized horseshoe base measure. This data-driven approach allows treatments to cluster based on their effects while formally propagating uncertainty about the clustering structure itself. We extend this method to incorporate baseline risk meta-regression, enabling clustering even under heterogeneity, and demonstrate implementation using standard MCMC software. We apply the method to case studies in rheumatology and pain and find adjusting for baseline risk heterogeneity can substantially change which treatments are clustered together, highlighting the importance of methods to allow for meta-regression.

## 2. Highlights

- **What is already known**

    o Network meta-analysis requires defining treatment nodes, but lumping similar treatments or splitting minor variants is often subjective and impacts results.

    o Heterogeneity across studies further complicates node definition and NMA interpretation.

    o Data-driven clustering methods exist but may be difficult to implement, ignore heterogeneity during the clustering process, or lack comprehensive uncertainty quantification regarding the nodes themselves.

- **What is new**

    o This paper introduces an adaptable Bayesian non-parametric method using a Dirichlet Process with a regularized horseshoe prior for NMA node clustering.

    o The approach allows data-driven clustering with full uncertainty propagation and is implemented using generic MCMC software (e.g., JAGS) for accessibility.

- o   We extend existing methods to incorporate meta-regression (demonstrated with baseline risk) to allow clustering while simultaneously accounting for effect modification by study-level covariates.

- o   The framework explicitly allows for incorporating prior domain knowledge about potential treatment groupings.

- **Potential impact for Research Synthesis Methods readers**

    - o   Provides readers with a practical, statistically rigorous, and flexible tool to address the common and challenging lumping/splitting dilemma in complex NMAs.

    - o   Improves the robustness and transparency of NMA findings by formally modeling and quantifying uncertainty related to treatment node definition.

    - o   Demonstrates the interaction between heterogeneity adjustment and identifying treatment similarity, leading to more nuanced and potentially less biased evidence synthesis.

# 3. Introduction

Network meta-analysis (NMA) allows for the simultaneous comparison of multiple treatments when direct, head-to-head trials are unavailable and, in some cases, can offer improved precision when networks have a mixture of direct and indirect evidence (Ades, Welton, Dias, Phillippo, & Caldwell, 2024). However, a critical challenge in NMA lies in defining treatment nodes, particularly when faced with multiple treatments or treatment variations that may be similar, but not identical. This issue, often framed as the dilemma between "lumping" similar treatments into a single node or "splitting" them into separate nodes, can profoundly impact the results and interpretation of an NMA (Xing & Lin, 2020).

The decision to lump or split treatments carries significant implications. Lumping distinct treatments may obscure differences in efficacy or safety, leading to overly broad and potentially misleading conclusions. Conversely, splitting highly similar treatments can increase the complexity of the network, leading to reduced precision and potentially spurious findings due to the increased number of parameters to be estimated. Barrientos, Page, & Lin (2024) illustrateillustrate this problem by showing that under simplifying normality assumptions, as the number of treatments increases in the meta-analysis, so do chances that at least one comparison is spuriously large. This combined multiplicity and magnitude problem has been discussed at length by Efthimiou & White (2020).

Xing & Lin (2020) explored the impact of different treatment classifications in a series of published NMAs, finding that the choice between lumping and splitting can substantively affect NMA results. They used the deviance information criterion (DIC) to evaluate model performance after lumping similar treatments and demonstrated that lumping can, in some cases, improve model fit. However, they also found that lumping can cause noticeable changes in effect estimates for many treatment comparisons.

While most lumping and splitting decisions approaches are qualitative in nature and are recommended to follow structured process combining clinical and methodological experts (Dhippayom, Saldanha, Chaiyakunapruk, & Devine, 2022), a few data-driven solutions have been proposed. Barrientos et al. (2024) introduced a Bayesian non-parametric approach that allows for formal consideration of uncertainty in treatment classification. Their approach uses a Dirichlet process prior with a spike-and-slab base measure to model treatment effects, allowing for a positive probability that two treatments have the same effect. More recently, Kong, Daly, & Béliveau (2024) demonstrate a frequentist approach to lumping based on applying the generalized fused lasso (GFL) to NMA in a two-step procedure. This method penalizes all pairwise differences in treatment effects, effectively pooling treatments that are not sufficiently different.

While a Bayesian approach is attractive for its ability to propagate all relevant uncertainty, the approach of Barrientos et al. (2024)relies on custom Markov Chain Monte Carlo (MCMC) samplers that can hinder its widespread adoption. Integrating this approach into existing, widely used generic sampler code, such as that provided in the NICE Decision Support Unit's (DSU) Technical Support Document 2 (TSD2) (Dias, Welton, Sutton, & Ades, 2011), would significantly enhance its accessibility and practical utility. Second, the use of a spike-and-slab prior can lead to challenges in interpreting the resulting models, particularly if the focus is placed on individual model probabilities instead of the broader goal of regularization and dimension reduction (Barrientos et al., 2024).

Third, both the approaches described by Barrientos et al. (2024) and Kong et al. (2024) primarily focus on pooling treatments based on their observed overall effects. However, heterogeneity in treatment effects may exist across studies due to variations in patient populations, study designs, or other factors. In particular, variations in baseline risk have been shown to be important effect modifiers in a variety of settings. Cameron et al. (2018)C et al illustrated how failing to adjust for differences in baseline risk in an NMA of interventions for psoriasis led to biased estimates of treatment effects and altered clinical interpretations. They demonstrated that baseline risk, which can serve as a proxy for multiple observed and unobserved patient and study characteristics, can significantly impact the magnitude and even direction of treatment effects. Existing methods do not adequately address the possibility of clustering treatments while simultaneously accounting for heterogeneity through meta-regression, particularly when adjusting for baseline risk and its associated uncertainty. While extensions to the fused lasso approach to incorporate meta-regression are possible, a Bayesian approach would allow for more flexibility including the ability to accurately model interactions with baseline risk.

Finally, while data-driven approaches are valuable, incorporating domain-specific knowledge into the clustering process remains an important challenge. In clinical settings with a large number of treatments, experts may possess prior knowledge about treatment similarities based on mechanisms of action, drug classes, or other relevant factors. Integrating this knowledge into the clustering process could help reduce the search space and lead to a more clinically plausible group of clusters. Conversely, in situations with limited prior knowledge, researchers may prefer a more exploratory approach, with fewer constraints on the number of possible clusters. Thus, a flexible framework that allows for incorporating varying degrees of prior knowledge into the clustering process is needed.

To address these limitations, this paper adapts and extends the Bayesian non-parametric approach of Barrientos et al. (2024). We propose replacing the spike-and-slab portion of the model with a regularized horseshoe prior (Piironen & Vehtari, 2017).as described by This approach achieves a similar goal of inducing sparsity and allowing for the potential pooling of treatments. However, regularized horseshoe priors maintain exclusive support over the parameter space, and can be more readily implemented using existing generic sampler code (e.g., JAGS) than priors used by Barrientos et al. (2024). We also frame the treatment effect estimation in terms of regularization rather than introducing potentially confusing discrete mixture models. Furthermore, we extend the approach to allow for clustering under potential treatment heterogeneity by incorporating meta-regression on baseline risk, while appropriately considering uncertainty in its estimation. The decision to lump or split treatments should be guided by clinical and methodological considerations, ensuring that the analysis accurately reflects the underlying treatment effects and heterogeneity. Balancing these approaches is essential for producing reliable and clinically relevant NMA results, and our methods will help facilitate this by allowing incorporation of domain knowledge into the clustering process, either by limiting the number of clusters or by specifying priors that favor more or less clustering. We illustrate the potential impact of these methodological advancements through case studies in NMAs for treatments for rheumatology and pain

## 4. Methods

### 4.1. Standard NMA Model

Our analysis builds upon the standard NMA model framework outlined in the NICE Decision Support Unit's Technical Support Document 2 (TSD2) (Dias, Welton, et al., 2011). Consider a binomial pairwise meta-analysis for simplicity:

$$\begin{aligned} r_{ik} &\sim \text{Binomial}(p_{ik}, n_{ik}) \\ \text{logit}(p_{ik}) &= \mu_i + \delta_{i,1k} I(k \neq 1) \\ \delta_{i,1k} &\sim N(d_{1k}, \tau^2) \end{aligned}$$

where $i$ and $k$ index studies and arms respectively, $r_{ik}$ is the number of patients with an event in $n_{ik}$ trials and is realized from a binomial distribution with probability $p_{ik}$ modeled using a logit link. The baseline risk for the trial is represented by $\mu_i$, and under the usual random effects assumption, the trial specific treatment effect is represented by $\delta_{i,1k}$ which is typically assumed to be sampled from a normal distribution with shared between-trial heterogeneity $\tau$.

### 4.2. Dirichlet Process Prior for Clustering

To account for potential clustering of treatment effects, we introduce a Dirichlet process (DP) prior distribution. The DP prior allows us to group treatments with similar effects into clusters, enabling a more nuanced understanding of the treatment landscape. In this implementation, the variable $cluster[k]$ indicates the cluster assignment for each treatment. Thecluster-specific treatment effects are specified as $\theta_h$ where $h$ is an index that ranges from 1 to H, and where H is a specified upper bound on the number of allowable clusters. Each $\theta_h$ is given a regularized horseshoe prior distribution, which is described in a following section. Following the notation of Barrientos et al. (2024), we can modify the standard model above to include a DP prior as follows:

$$\begin{aligned} p &\sim \text{DP}(\alpha, P_0) \\ cluster_t &\sim \text{Categorical}(\mathbf{p}) \\ \delta_{i,1k} &\sim N\left(\theta_{cluster_{t_{i,k}}}, \sigma^2\right) \\ \text{where } P_0 &\equiv \mathcal{N}(0, \tau^2 \tilde{\lambda}_h^2) \\ \theta_h &\sim P_0 \\ \text{and } \tilde{\lambda}_h^2 &= \frac{c^2 \lambda_h^2}{c^2 + \tau^2 \lambda_h^2} \\ \lambda_h &\sim C^+(0,1) \\ \tau &\sim C^+(0, \tau_0) \end{aligned}$$

where $cluster_t$ indicates the cluster assignment for treatment $t$;, $\theta_h$ is the effect for cluster $h$;, p is a vector of cluster probabilities; $P_0$ is the base distribution for the DP prior, which in this case, is the regularized horseshoe distribution (described below); $\alpha$ is the concentration parameter of the DP; and $H$ is the number of clusters.

The DP prior assumes that the cluster probabilities **p** are drawn from a Dirichlet process with concentration parameter $\alpha$ and base distribution $P_0$. This is implemented using a stick-breaking construction in the corresponding JAGS code. The effect for each treatment, $\delta_{i,k}$, is then drawn from a normal distribution centered around the effect of its assigned cluster, $\theta_{cluster[t[i,k]]}$, with variance $\sigma^2$. The variable $t[i,k]$ indicates the treatment used in each arm of the data.

To address potential issues related to sensitivity to the baseline treatment and to improve the interpretability of our results, we deviate from the spike and slab approach described by Barrientos et al. (2024) and instead employ a regularized horseshoe prior for the treatment effects. The regularized horseshoe prior can be viewed as a continuous alternative to the discrete spike and slab prior (Piironen & Vehtari, 2017). It is defined as a hierarchical prior with the following structure $\tau$ is a global shrinkage parameter that controls the overall sparsity of the model, $\lambda_h$ is a local shrinkage parameter that allows individual effects to escape shrinkage, $c$ is a slab width parameter that determines the degree of regularization for large effects and $C^+(0,1)$ is the standard half-Cauchy distribution. This prior has several advantages.. First, it provides a continuous relaxation of the spike-and-slab prior, making it more amenable to MCMC sampling. Next, it allows for separate control over sparsity (via $\tau$) and the regularization of large effects (via $c$).). Additionally, it avoids the need for non-standard priors such as the DNP by using a regularized structure to ensure separation between the "spike" and "slab" components. Last. Last, it shrinks all effects towards zero to some extent which helps to prevent issues with multiplicity. The choice of the slab-width parameter c determines the degree of regularization applied to larger effects allowed to escape shrinkage. We follow recommendations from Piironen & Vehtari (2017) by setting a prior on c such that the slab is t distributed with 1 degree of freedom and a standard deviation of 1 which when combined with their recommended priors on $\tau$ leads to a marginal 95% prior predictive of credible interval of approximately -5 to 5.

## 4.3. Extension to Baseline Risk Meta-Regression

The model can be further extended to incorporate baseline risk as a covariate, following the approach described in NICE DSU Technical Support Document 3 (TSD3) (Dias, Welton, et al., 2011). This allows us to account for potential differences in treatment effects across studies due to variations in the underlying risk of the outcome in the control group. The model, assuming a shared interaction term, extends the model above by modifying the linear predictor term for each arm within a study as follows.

$$\text{logit}(p_{ik}) = \mu_i + \delta_{i,1k}I(k \neq 1) + (\beta_{t[i,k]} - \beta_{t[i,1]})(x_i - \bar{x})$$

where $x_i$ is the baseline risk for study $i$, centered around the mean baseline risk $\bar{x}$ and $\beta_{i,k}$ is the interaction term for treatment $k$ in study $i$. In this model, the interaction term $\beta$ represents the change in the log-odds ratio of the treatment effect for a one-unit increase in the baseline risk.

## 4.4. Case Study Data

### 4.4.1. Certolizumab

This dataset originates from a review conducted in support of a NICE Single Technology Appraisal (TSD3) (Dias, Sutton, Welton, & Ades, 2011) evaluating Certolizumab Pegol (CZP) for rheumatoid arthritis (RA) in patients who had failed on disease-modifying anti-rheumatic drugs (DMARDs), including Methotrexate (MTX). It comprises data from 12 MTX-controlled randomized controlled trials (RCTs) comparing seven different treatments: Placebo plus MTX, CZP plus MTX, Adalimumab plus MTX, Etanercept plus MTX, Infliximab plus MTX, Rituximab plus MTX, and Tocilizumab plus MTX. The primary outcome extracted from these trials was the number of patients achieving ACR50 response at 6 months (or 3 months if 6-month data was unavailable) within each treatment arm.

### 4.4.2. Pain

This dataset was used to illustrate baseline risk adjustment NMA consists of data from 56 RCTs, yielding 116 data points (Achana et al., 2013). The trials investigated the effectiveness of three non-opioid analgesics – paracetamol, non-steroidal anti-inflammatory drugs (NSAIDs), and cyclooxygenase 2 (COX-2) inhibitors – compared to placebo in reducing morphine consumption in adults following major surgery. The outcome measured was the amount of morphine consumed in milligrams over a 24-hour period, measured as a continuous variable. For each treatment arm within each trial, the dataset provides the number of patients, the mean 24-hour morphine consumption, and its standard deviation. Notably, there is substantial variability in baseline morphine consumption in the placebo arms across the trials. Two trials lacked a placebo arm, and these were addressed in the original analysis through imputation and in the current analysis using the exchangeability assumption within a network meta-analysis framework. In the current analysis,the placebo arms were excluded from those trials.

## 4.5. Implementation Details

The analysis was conducted using JAGS. We assessed convergence using trace plots and the Gelman-Rubin statistic where values less than 1.05 were considered to have converged (Gelman & Rubin, 1992). Model code and data listings for all analyses are presented in the appendix. Network structure is shown in network diagrams where the width of lines connecting nodes indicates the number of comparisons between treatments, and the size of the nodes is proportional to the sample size (binary outcomes) and the sum of inverse variance (continuous outcomes when only se is reported for studies). We present results in terms of odds ratios (ORs) relative to a reference treatment for binary outcomes and mean differences (MDs) for continuous outcomes along with their 95% credible intervals. Comparisons between treatments are summarized using league tables where higher ranking treatments are ordered to the top left corner and lower ranking treatments to the bottom right.  Clustering models are additionally summarized by a heatmap showing the probability that each treatment is clustered with every other treatment, and the modal network with the proportion of times it was selected. Models are compared in terms of absolute fit via total residual deviance and expected out of sample performance using deviance information criteria (DIC) where a difference of 3-5 may indicate preference for a more complex model (Dias, Welton, et al. (2011)).

It is worth noting that unlike Barrientos et al. (2024), the approach described by the current method does not use conditional credible intervals. This is because the use of the regularized horseshoe prior avoids issues with interpretation in discrete mixtures, allowing us to focus on the marginal effects estimated.

## 4.6. Informing the Prior Distribution on number of clusters through Domain Knowledge

The proposed methods allow for the incorporation of domain knowledge through the specification of prior distributions for the model parameters. For instance, we can modify the parameters of the DP prior to reflect our beliefs about the number of clusters. Specifically, we can limit the maximum number of clusters $H$ to a value that is considered reasonable for decision-making purposes. This can be achieved by adjusting the prior for the concentration parameter $\alpha$ of the DP. We can check the implied priors on the number of clusters through prior predictive checks. For these case studies, we use $H$ equal to the number of treatments and $\alpha$ of 1. This translates into a prior of 2 (95% CrI 1 to 4) for the certolizumab example and 2 (1 to 3) for the pain example. Remaining prior distributions for all parameters are described in the appendices.

## 4.7. Code availability

All code to replicate analyses is available at https://github.com/timdisher/dpNMA

# 5. Results

## 5.1. Certolizumab Case Study

**Figure 1.** *Network diagram for certolizumab case study*

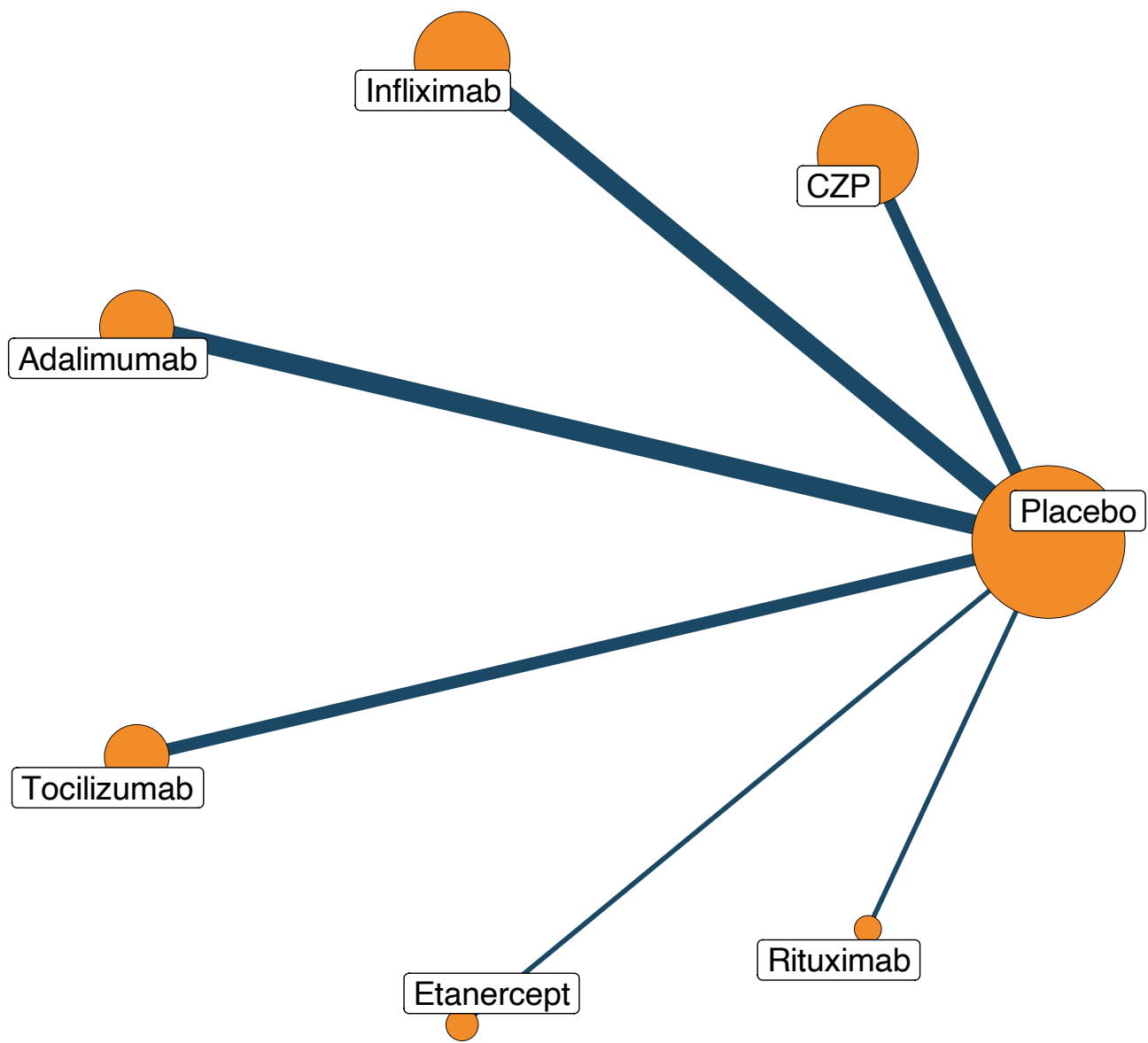

Node size indicates number of patients and line thickness indicates number of trials.
Abbreviations: CZP - certolizumab pegol

In the certolizumab network (Figure *4*), the Deviance Information Criterion (DIC) values (Table *1* ) indicated improved model fit when incorporating the Dirichlet process prior (DP BLR: 152, DP Unadjusted: 155) compared to the standard models (BLR: 154, Unadjusted: 156). This suggests that the data support the possibility of some treatment effects being similar enough to be considered as belonging to the same cluster. The coefficient for baseline risk was similar for the DP and standard BLR models, with credible intervals excluding the null in both cases. Models adjusting for baseline risk also had better absolute and relative model fit in addition to notably lower between-trial heterogeneity (BLR: 0.19, DP BLR: 0.25, Unadjusted: 0.75, DP Unadjusted: 0.66). Based on these findings, the DP BLR model may be considered preferred given it is more parsimonious than the standard BLR model.

**Table 1.** *Model fit statistics for certolizumab case study*

| Model | B | Total residual deviance | Between-trial SD | DIC |
| --- | --- | --- | --- | --- |
| DP BLR | -0.95 (-1.06 to -0.71) | 24.61 | 0.25 (0 to 0.55) | 152 |
| BLR | -0.99 (-1.09 to -0.77) | 24.11 | 0.19 (0 to 0.54) | 154 |
| Unadjusted | | 27.44 | 0.72 (0.01 to 1.61) | 155 |
| DP Unadjusted | | 27.65 | 0.67 (0.15 to 1.43) | 155 |

B=beta; BLR= baseline risk; DIC= Deviance Information Criterion; DP= Dirichlet process; SD= standard deviation

The league table of pairwise odds ratios (Figure *2*, Panel A) from the standard BLR model showed that all active treatments except rituximab were significantly superior to placebo. The DP BLR model (Figure *2*, Panel B), shows general agreement with those conclusions, but estimates a high probability that all actives treatment other than rituximab should be combined, and has slightly less certainty that therapies have a high probability of being better than rituximab. This can be seen in the heatmap of DP clustering in the BLR adjusted model (Figure *3* Panel B), where rituximab is combined with active therapies ~ 15% of the time.

**Figure 2.** *League tables for certolizumab case study*

## A

| | | | | | | |
|---|---|---|---|---|---|---|
| **Tocilizumab** | | | | | | |
| 1.09 (0.58 to 2.04) | **Adalimumab** | | | | | |
| 1.29 (0.49 to 3.54) | 1.19 (0.46 to 2.97) | **Etanercept** | | | | |
| 1.55 (0.77 to 3.08) | 1.42 (0.79 to 2.66) | 1.21 (0.47 to 2.93) | **CZP** | | | |
| 1.73 (0.89 to 3.35) | 1.59 (0.89 to 2.87) | 1.34 (0.53 to 3.32) | 1.12 (0.58 to 1.98) | **Infliximab** | | |
| 6.46 (2.09 to 23.52) | 5.94 (1.97 to 20.89) | 4.99 (1.29 to 21.82) | 4.16 (1.33 to 14.61) | 3.72 (1.23 to 13.11) | **Rituximab** | |
| 9.54 (5.92 to 16.19) | 8.75 (6.02 to 13.51) | 7.39 (3.30 to 17.20) | 6.15 (3.85 to 9.47) | 5.50 (3.64 to 8.49) | 1.48 (0.45 to 4.15) | **Placebo** |

## B

| | | | | | | |
|---|---|---|---|---|---|---|
| **Tocilizumab** | | | | | | |
| 1.00 (1.00 to 1.17) | **Adalimumab** | | | | | |
| 1.00 (1.00 to 1.61) | 1.00 (1.00 to 1.57) | **Etanercept** | | | | |
| 1.00 (1.00 to 1.64) | 1.00 (1.00 to 1.62) | 1.00 (1.00 to 1.62) | **CZP** | | | |
| 1.00 (1.00 to 1.69) | 1.00 (1.00 to 1.67) | 1.00 (1.00 to 1.64) | 1.00 (1.00 to 1.47) | **Infliximab** | | |
| 5.68 (1.00 to 11.99) | 5.68 (1.00 to 11.92) | 5.55 (1.00 to 11.90) | 5.41 (1.00 to 11.56) | 5.32 (1.00 to 11.49) | **Rituximab** | |
| 7.23 (5.57 to 10.10) | 7.23 (5.58 to 9.95) | 7.12 (5.20 to 9.69) | 7.01 (5.20 to 9.30) | 6.94 (4.94 to 9.23) | 1.28 (0.60 to 8.12) | **Placebo** |

**Treatment in column has 97.5% or greater probability of being better than treatment in row** — No / Yes

A: Baseline risk adjusted standard model; B: Baseline risk adjusted DP model. Abbreviations: CZP - certolizumab pegol

Comparing the lumping of treatments in the unadjusted and adjusted DP models shows how heterogeneity can influence lumping decisions. In both the unadjusted and adjustedj DP models (Figure *3* Panel,s A and B), active therapies other than rituximab showed similarly high probability of being grouped into the same cluster. In the unadjusted model, however, rituximab has an approximately 50% probability of being included in this cluster whereas that probability dropped to approximately 15% when heterogeneity in baseline risk was accounted for. This difference was also found in the modal network, which combined all therapies in the unadjusted model but with large uncertainty (41%) versus combining all therapies except rituximab after adjustment with increased confidence (64%).

**Figure 3.** *Heatmap for treatment clustering in certolizumab case study*

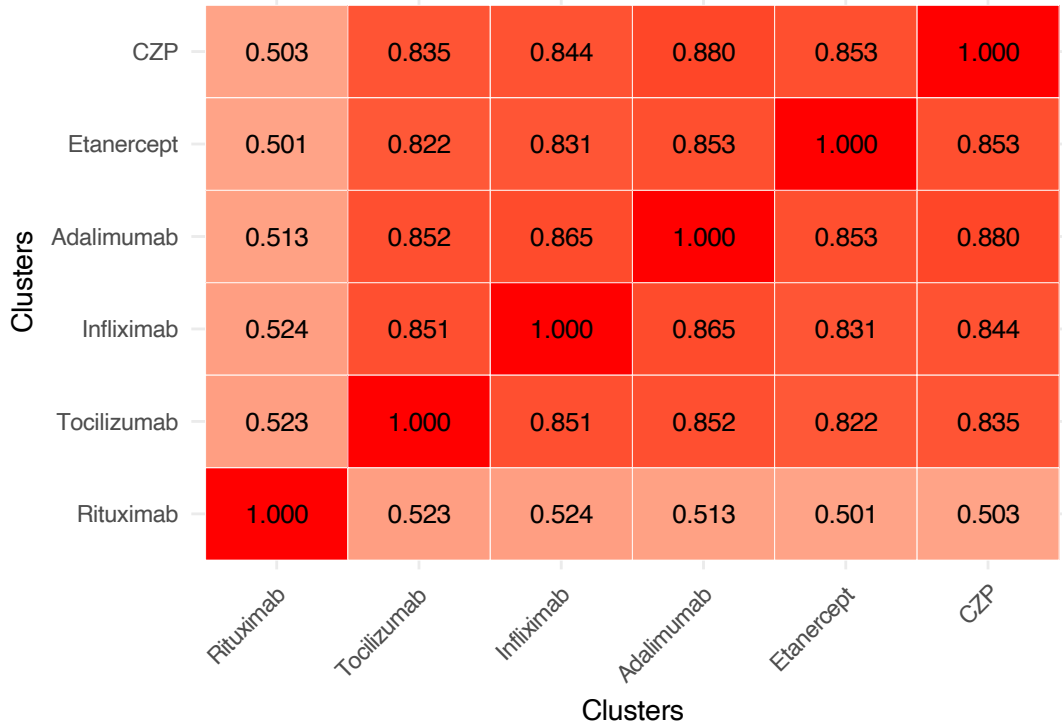
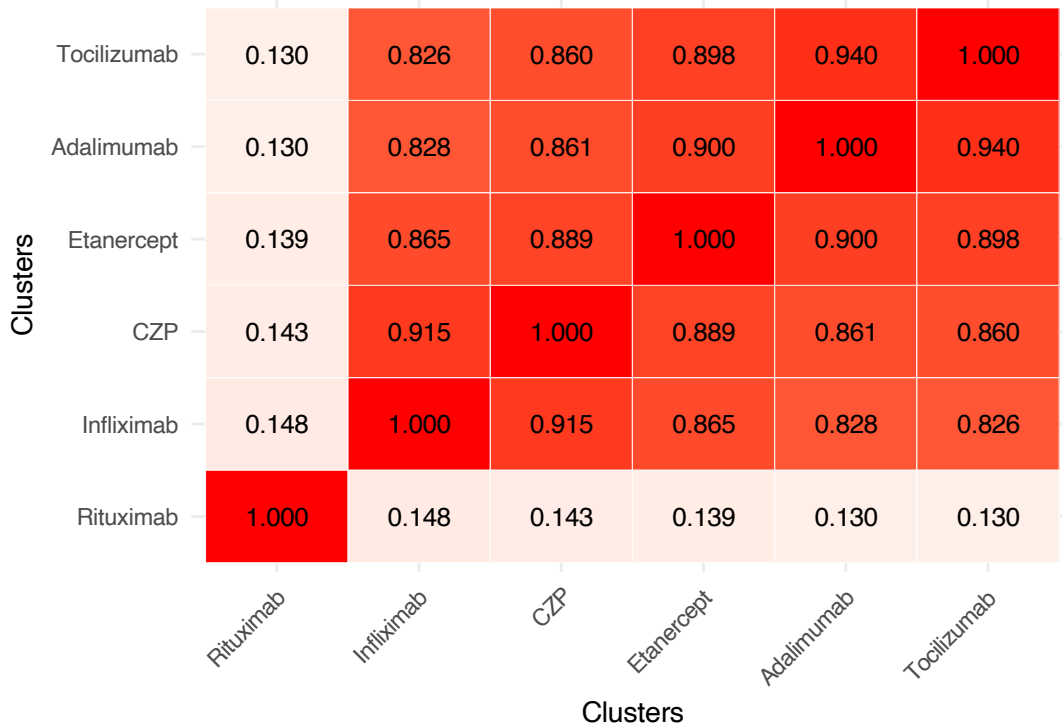

A: Unadjusted DP model; B: Baseline risk adjusted DP model.
Abbreviations: CZP - certolizumab pegol

## 5.2. Pain Case Study

We further evaluated the performance of the Bayesian Dirichlet process model with a regularized horseshoe prior using a network meta-analysis of post-operative pain management strategies, employing the dataset from Achana et al. This dataset included 56 trials assessing the impact of three active treatments – paracetamol, NSAIDs, and COX-2 inhibitors – compared to placebo on morphine consumption (Figure *4*.

**Figure 4.** *Network diagram for pain case study*

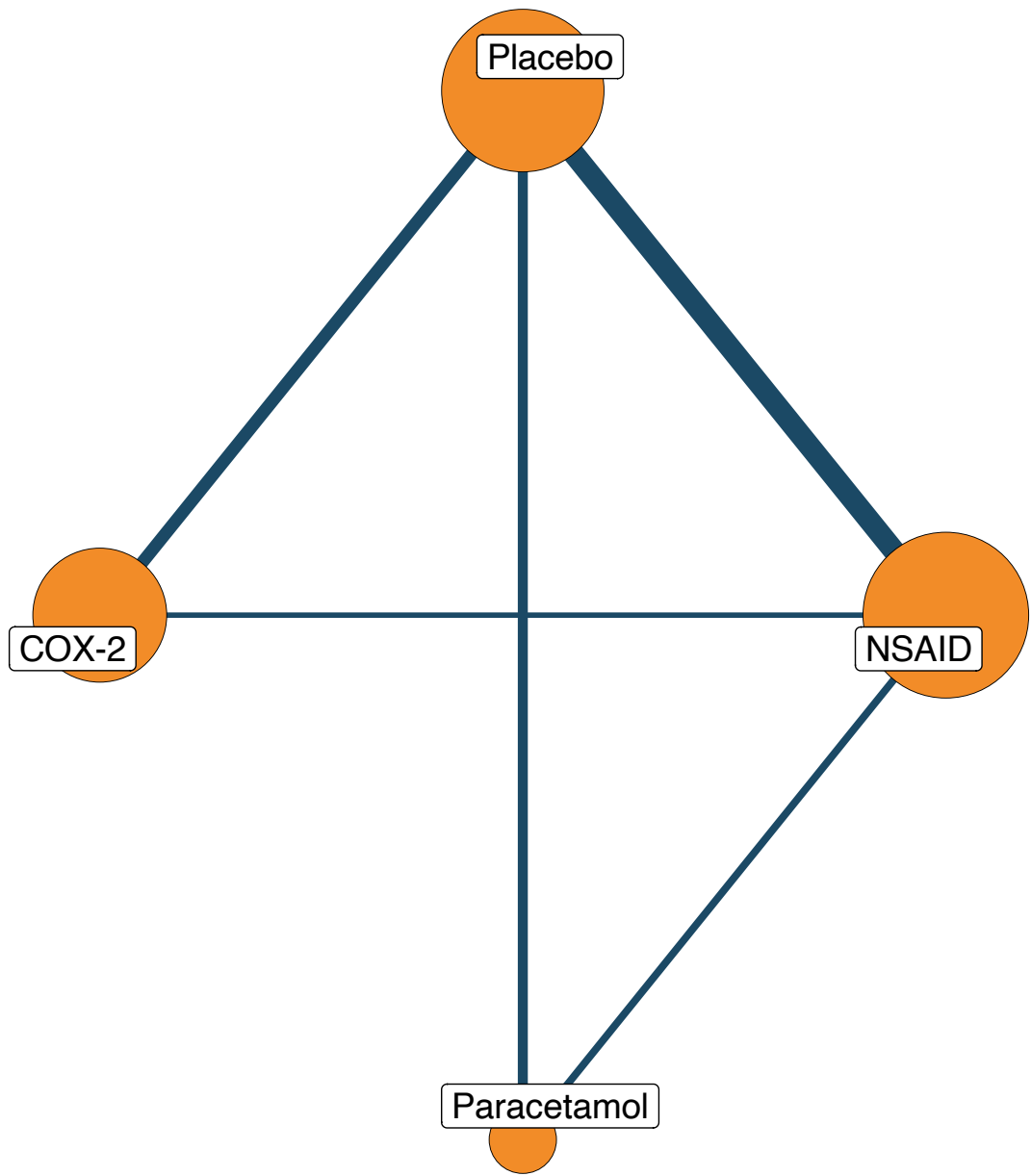

Node size indicates number of patients and line thickness indicates number of trials.
Abbreviations: COX-2 - cyclooxygenase-2; NSAID - non-steroidal anti inflammatory

The Deviance Information Criterion (DIC) values (Table *2*) indicated that models incorporating baseline risk adjustment (BLR: 154, DP BLR: 152) demonstrated a better fit to the data compared to the unadjusted models (Unadjusted: 156, DP Unadjusted: 154). While this difference in DIC between the adjusted and unadjusted models is small, the reduction in between-trial heterogeneity is very large, and betas for the baseline risk interaction term are precisely estimated with high probability that coefficients are less than 0 which is consistent with the BLR models being preferred. Within the BLR models, selection of a preferred model is more difficult since between-trial heterogeneity is larger but more precisely estimated in the DP BLR model and DIC is lower by two points.

**Table 2.** *Model fit statistics for pain case study*

| Model | B | Total residual deviance | Between-trial SD | DIC |
| --- | --- | --- | --- | --- |
| DP BLR | -0.95 (-1.06 to -0.71) | 24.61 | 0.25 (0 to 0.55) | 152 |
| BLR | -0.99 (-1.09 to -0.77) | 24.11 | 0.19 (0 to 0.54) | 154 |
| Unadjusted | | 27.44 | 0.72 (0.01 to 1.61) | 155 |
| DP Unadjusted | | 27.65 | 0.67 (0.15 to 1.43) | 155 |

B=beta; BLR= baseline risk; DIC= Deviance Information Criterion; DP= Dirichlet process; SD= standard deviation

Looking at point estimates of treatment comparisons, the league table summarizing the differences in mean morphine consumption (Figure *5*, Panel A) from the standard BLR model and DP BLR panel (Panel B) provide similar conclusions and estimates compared to placebo, but lumping of active therapies leads to large differences in the point estimates compared to the standard model. Estimates from the lumped model are more precise, with the decrease in variance consistent with a 1.62 to 4.24 times increase in effective sample size.

**Figure 5.** *League tables for pain case study*

**A**

| COX-2 | | | |
|---|---|---|---|
| -1.22 (-3.89 to 1.43) | NSAID | | |
| -2.34 (-6.09 to 1.32) | -1.12 (-4.54 to 2.24) | Paracetamol | |
| -13.40 (-15.64 to -11.22) | -12.16 (-13.91 to -10.51) | -11.04 (-14.14 to -7.85) | Placebo |

**B**

| COX-2 | | | |
|---|---|---|---|
| 0.00 (0.00 to 0.00) | NSAID | | |
| 0.00 (-0.05 to 0.02) | 0.00 (0.00 to 0.00) | Paracetamol | |
| -12.36 (-13.74 to -10.99) | -12.33 (-13.67 to -10.98) | -12.31 (-13.77 to -10.86) | Placebo |

**Treatment in column has 97.5% or greater probability of being better than treatment in row** — No / Yes

A: Baseline risk adjusted standard model; B: Baseline risk adjusted DP model.
Abbreviations: COX-2 - cyclooxygenase-2; NSAID - non-steroidal anti inflammatory

As in the certolizumab example, baseline risk adjustment led to large changes in the pairwise clustering of treatments (Figure *6*). In the unadjusted model (Panel A), paracetamol had an approximately 75% chance of being combined with COX-2 and NSAIDs but the adjusted model found that all therapies had a greater than 95% probability of being combined. In both cases, the modal network combined all active therapies but was more certain after adjustment (95%) than in the unadjusted model (74%). Considering the high confidence in lumping, greater model parsimony and general alignment in conclusions, the DP BLR model may considered preferred.

**Figure 6.** *Heatmap for treatment clustering for pain case study*

A

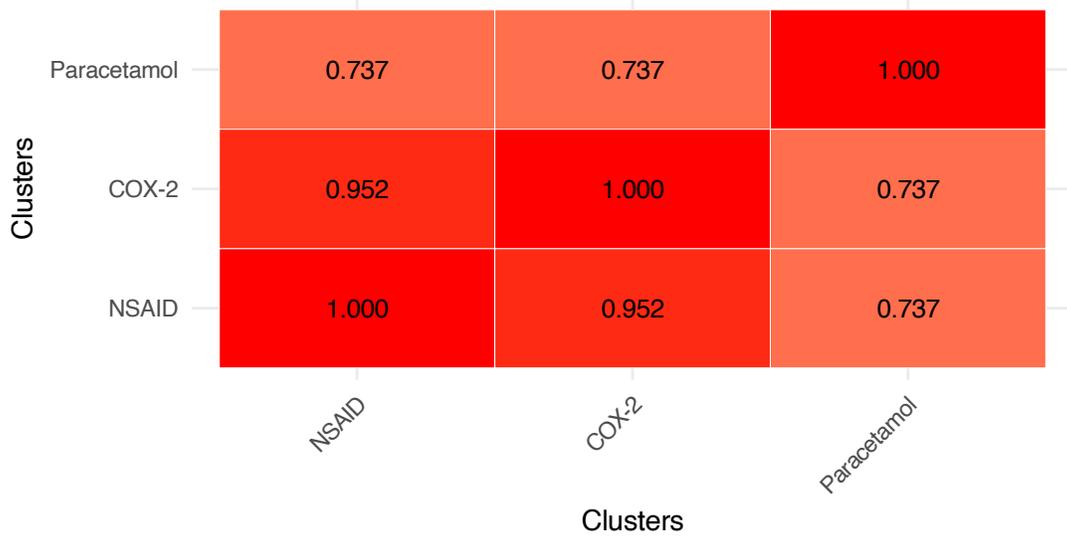

B

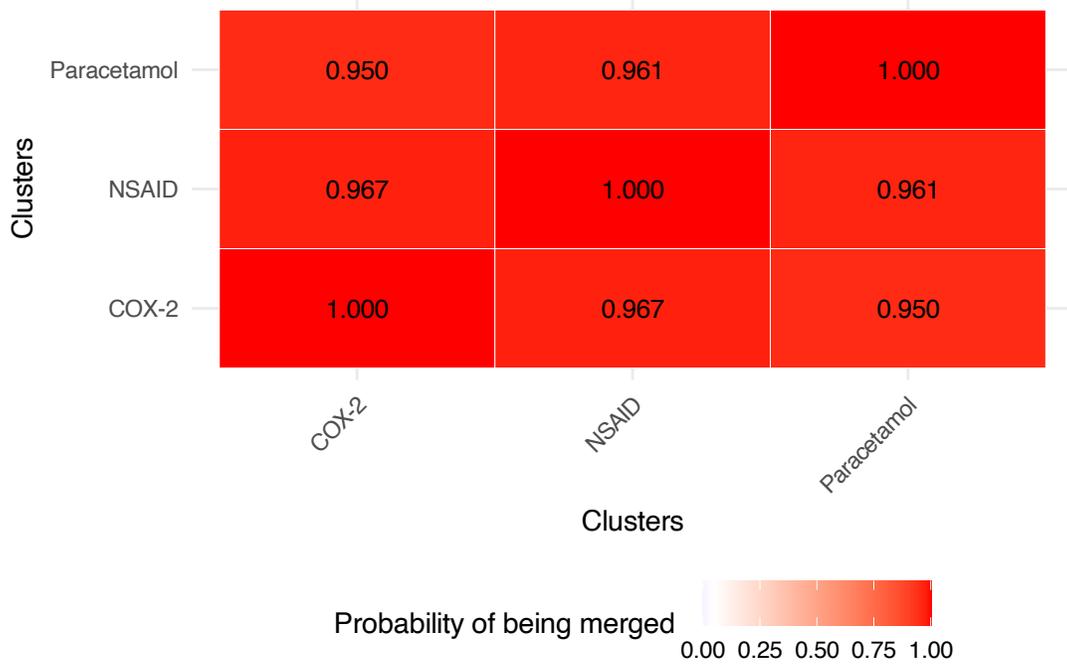

A: Unadjusted DP model; B: Baseline risk adjusted DP model.
Abbreviations: COX-2 - cyclooxygenase-2; NSAID - non-steroidal anti inflammatory

## 6. Discussion

This study extends the Bayesian nonparametric approach of Barrientos et al. (2024) for addressing the complexities of treatment node definition in network meta-analysis (NMA).

Our work isis novel in several key ways: by adapting this approach for use with generic samplers, implementing a simpler spike-and-slab alternative using the regularized horseshoe prior, extending the methodology to accommodate clustering under heterogeneity, and incorporating expert opinion on the number of clusters.

Lumping and splitting are crucial considerations in NMAnetwork meta-analysis (NMA) when merging nodes, such as treatments within a class of drugs. Lumping involves combining treatments that are assumed to have similar effects, which can simplify the analysis and increase statistical power by pooling data. This approach is particularly useful when individual treatments are expected to perform similarly, allowing for more robust conclusions. On the other hand, splitting involves treating each treatment as a distinct node, which can provide more detailed insights but may reduce statistical power due to smaller sample sizes. Balancing these approaches is essential for producing reliable and clinically relevant NMA results, and our methods will help facilitate this.

Our primary finding is that the choice between which treatments to lump can differ between adjusted and unadjusted models, suggesting that heterogeneity may disguise similarity in treatments or create spurious clusters. Lumping decisions also had large effects on the effective sample size of comparisons which may lead to substantively different conclusions in some networks. This builds upon the work of Xing & Lin (2020), who demonstrated that different treatment classifications can meaningfully affect NMA results, and extends the work of Cameron et al. (2018), who highlighted the importance of adjusting for baseline risk in NMA, by allowing for clustering in the presence of baseline risk heterogeneity. This is particularly relevant given that network meta-regression is typically limited in the number of variables that can feasibly be adjusted and baseline risk can serve as a proxy for multiple observed and unobserved patient and study characteristics, as noted by Cameron et al. (2018).

Decisions to lump nodes are typically driven primarily by clinical understanding including pathophysiology, drug targets, doses, routes, and other considerations (Dhippayom et al., 2022). Purely statistical decisions are usually avoided, since clinical criteria itself may be considered during decision making. For example, it may be possible that a less efficacious drug from a well-established class of therapies can be favored as a first line treatment over newer therapies where there is less certainty in the efficacy/safety profiles in prolonged real-world use. The Bayesian implementation presented here is perhaps a better conceptual fit than two-stage approaches in these case, since network diagrams and node labels are retained distinct and the clustering process can be considered more of a regularizing/simplifying step instead of being used to drive node formation itself. This method also fits well within and strengthens the objectivity of the GRADE minimally contextualized difference approach, which involves a manual and somewhat arbitrary grouping of therapies together in terms of efficacy/safety profiles (Brignardello-Petersen et al., 2020). The method described here may further be of value in guiding initial exploration and pressure-testing of lumping decisions. This may be particularly helpful in cases where "classes" of therapies are provided less prior weight or many doses/combinations of therapies make lumping decisions more difficult. Our approach also offers the first opportunity we are aware of to incorporate domain knowledge into the lumping process by limiting the number of clusters or specifying priors that favor more or less clustering. This addresses the challenge of integrating expert opinion with data-driven approaches, allowing researchers to tailor the analysis to their specific context.

We also address several limitations of the original Barrientos et al. (2024) approach. Their reliance on custom MCMC samplers presented a barrier to wider adoption. We adapted their method to work with generic samplers, such as those provided in the NICE DSU TSD2 (Dias, Welton, et al. (2011), thereby enhancing its accessibility and practical utility. Additionally, we replaced their spike-and-slab base measure, which can suffer from mixing and interpretability problems, with a regularized horseshoe prior Piironen & Vehtari (2017). This continuous spike-and-slab alternative not only improves MCMC mixing but also enhances interpretability by framing the problem in terms of regularization rather than discrete mixture models. The regularized horseshoe prior shrinks all effects to some extent which may further help to reduce issues of multiplicitymultiplicity in NMA. The work of Efthimiou & White (2020) highlighted the problems with multiple comparisons in NMA, especially in regard to false positives. Our work helps address that issue by providing a mechanism to control for false positive results by simultaneously shrinking the dimensionality of the network and providing a safeguard against spuriously large effects from smaller trials.

Our approach provides a Bayesian alternative to the generalized fused lasso (GFL) problem presented by Kong et al. (2024). While the GFL offers a frequentist approach to treatment pooling, our Bayesian framework allows for the incorporation of prior information and the full propagation of uncertainty through a one-step model that itself can be expanded to accommodateaccommodate many different likelihoods or other model needs. The Bayesian paradigm is particularly well-suited to handling the complexities of NMA, where uncertainty quantification is crucial for informed decision

However, our approach also has limitations. Compared to the generalized fused lasso, our current model creates probabilistic pairings, while some decision-makers may prefer deterministic lumping for strict dimension reduction. While we showed two ways to achieve deterministic lumping (identifying the modal network and exploration of the pairwise matrix), future work should compare this to the fused lasso approach in terms of computational efficiency and decision-making utility. We also found that use of both the pairwise matrix and modal network was complementary, since the latter allowed for more precise understanding of where increased certainty in the modal network was informed from. Additionally, creating a-step procedure, where we first identify clusters and then fit a standard NMA model, may not fully propagate uncertainty from the clustering step to the final treatment effect estimates. Although previous simulations have not shown this to be a major issue Kong et al. (2024), it would be valuable to compare estimates across a broad range of case studies and assess the potential miscalibration of either model.

Opportunities for future research may include extensions to allow for explicit prior information on which therapies should be combined. This could perhaps be implemented via hierarchical processes where therapies within, for example, the same class of drugs may be more likely to be clustered together (Liu, Wu, Jiang, Boot, & Rozen, 2023). This would address a major limitation of the current approach for users that would like to use the method within a traditional node definition process while recognizing the uncertainty in aspects of those decisions. The current approach also treats lumping decisions across multiple outcomes as distinct, which may be problematic if patterns of lumping across multiple efficacy and safety outcomes lacks face validity. Formally addressing this issue would require multivariate methods, but may be of interest for researchers focused on creating groups of therapies with similar benefit/safety profiles.

## 7. Conclusion

In conclusion, this study provides a robust and flexible Bayesian framework for addressing the lumping/splitting dilemma in NMA. By incorporating a Dirichlet process prior with a regularized horseshoe base measure, extending the model to accommodate meta-regression on baseline risk, and allowing for the incorporation of expert opinion, our approach offers a more nuanced and interpretable way to analyze complex treatment networks. These methodological advancements have the potential to improve the accuracy and relevance of NMA results for clinical decision-making. Our findings also underscore the importance of carefully considering heterogeneity when defining treatment nodes and highlight the need for further research on deterministic lumping and uncertainty propagation in Bayesian NMA models.

# 9. Appendix

## 9.1. Certolizumab case study

### 9.1.1. Data

```
## list(t = structure(c(1L, 1L, 1L, 1L, 1L, 1L, 1L, 1L, 1L, 1L,
## 1L, 1L, 3L, 3L, 3L, 2L, 2L, 5L, 5L, 5L, 4L, 6L, 7L, 7L), dim = c(12L,
## 2L), dimnames = list(NULL, c("t_1", "t_2"))), na = c(2L, 2L,
## 2L, 2L, 2L, 2L, 2L, 2L, 2L, 2L, 2L, 2L), ns = 12L, nt = 7L, H = 7L,
##     nu = 1, s2 = 1, r = structure(c(9L, 19L, 5L, 15L, 4L, 33L,
##     22L, 0L, 1L, 5L, 14L, 22L, 28L, 81L, 37L, 146L, 80L, 110L,
##     61L, 15L, 23L, 5L, 26L, 90L), dim = c(12L, 2L), dimnames = list(
##         NULL, c("r_1", "r_2"))), n = structure(c(63L, 200L, 62L,
##     199L, 127L, 363L, 110L, 47L, 30L, 40L, 49L, 204L, 65L, 207L,
##     67L, 393L, 246L, 360L, 165L, 49L, 59L, 40L, 50L, 205L), dim = c(12L,
##     2L), dimnames = list(NULL, c("n_1", "n_2"))), mx = -2.41645197648634)
```

### 9.1.2. Model code

#### 9.1.2.1. Standard NMA

```
"model {
  for(i in 1:ns) {   # Loop through studies
    w[i,1] <- 0  # Adjustment for multi-arm trials is zero for control arm
    delta[i,1] <- 0   # Treatment effect is zero for control arm
    mu[i] ~ dnorm(0, .0001)   # Vague priors for all trial baselines
    for (k in 1:na[i]) {   # Loop through arms
      r[i,k] ~ dbin(p[i,k], n[i,k])   # Binomial likelihood
      logit(p[i,k]) <- mu[i] + delta[i,k]   # Model for linear predictor
      rhat[i,k] <- p[i,k] * n[i,k]   # Expected value of the numerators
      dev[i,k] <- 2 * (r[i,k] * (log(r[i,k]) - log(rhat[i,k]))   # Deviance contribution
        + (n[i,k] - r[i,k]) * (log(n[i,k] - r[i,k]) - log(n[i,k] - rhat[i,k])))
    }
    resdev[i] <- sum(dev[i,1:na[i]])   # Summed residual deviance contribution for this trial

    for (k in 2:na[i]) {   # Loop through arms
      delta[i,k] ~ dnorm(md[i,k],taud[i,k]) # trial-specific LOR distributions
      md[i,k] <- d[t[i,k]] - d[t[i,1]] + sw[i,k] # mean of treat effects distributions (with multi-arm trial correction)
      taud[i,k] <- tau_sd *2*(k-1)/k # precision of treat effects distributions (with multi-arm trial correction)
      w[i,k] <- (delta[i,k] - d[t[i,k]] + d[t[i,1]]) # adjustment for multi-arm RCTs
      sw[i,k] <- sum(w[i,1:(k-1)])/(k-1) # cumulative adjustment for multi-arm trials
```

```
    }
  }
  totresdev <- sum(resdev[])   # Total Residual Deviance
  d[1] <- 0   # Treatment effect zero for reference treatment

  for (k in 2:nt) {
    d[k] ~ dnorm(0, 0.001)
  }

   for (c in 1:(nt-1)) {
    for (k in (c+1):nt) {
      diff[c,k] <- (d[k]-d[c])
    }
  }
  sd ~ dunif(0, 2)
  tau_sd <- pow(sd, -2)

}
"
```

### 9.1.2.2. Standard NMA with meta-regression

```
model {
  for(i in 1:ns) {   # Loop through studies
    w[i,1] <- 0   # Adjustment for multi-arm trials is zero for control arm
    delta[i,1] <- 0   # Treatment effect is zero for control arm
    mu[i] ~ dnorm(0, .0001)   # Vague priors for all trial baselines

    for (k in 1:na[i]) {   # Loop through arms
      r[i,k] ~ dbin(p[i,k], n[i,k])   # Binomial likelihood
      logit(p[i,k]) <- mu[i] + delta[i,k] + (beta[t[i,k]]-beta[t[i,1]]) * (mu[i] - mx) # Model for linear predictor
      rhat[i,k] <- p[i,k] * n[i,k]   # Expected value of the numerators
      dev[i,k] <- 2 * (r[i,k] * (log(r[i,k]) - log(rhat[i,k]))   # Deviance contribution
        + (n[i,k] - r[i,k]) * (log(n[i,k] - r[i,k]) - log(n[i,k] - rhat[i,k])))
    }
    resdev[i] <- sum(dev[i,1:na[i]])   # Summed residual deviance contribution for this trial

    for (k in 2:na[i]) {   # Loop through arms
      delta[i,k] ~ dnorm(md[i,k],taud[i,k]) # trial-specific LOR distributions
      md[i,k] <- d[t[i,k]] - d[t[i,1]] + sw[i,k] # mean of treat effects distributions (with multi-arm trial correction)
      taud[i,k] <- tau_sd *2*(k-1)/k # precision of treat effects distributions (with multi-arm trial correction)
      w[i,k] <- (delta[i,k] - d[t[i,k]] + d[t[i,1]]) # adjustment for multi-arm RCTs
```

```
      sw[i,k] <- sum(w[i,1:(k-1)])/(k-1) # cumulative adjustment for multi-arm trials

    }
  }
  totresdev <- sum(resdev[])  # Total Residual Deviance
  d[1] <- 0  # Treatment effect zero for reference treatment
  beta[1] <- 0
  for (k in 2:nt) {
    d[k] ~ dnorm(0, 0.001)
    beta[k] <- B
  }

   for (c in 1:(nt-1)) {
    for (k in (c+1):nt) {
      diff[c,k] <- (d[k]-d[c])
    }
   }

  # vague priors for treatment effects
  B ~ dnorm(0,.001)
  sd ~ dunif(0, 2)
  tau_sd <- pow(sd, -2)

}
```

### 9.1.2.3. DP NMA

```
model {
  for(i in 1:ns) {   # Loop through studies
    w[i,1] <- 0  # Adjustment for multi-arm trials is zero for control arm
    delta[i,1] <- 0   # Treatment effect is zero for control arm
    mu[i] ~ dnorm(0, .001)   # Vague priors for all trial baselines

    for (k in 1:na[i]) {   # Loop through arms
      r[i,k] ~ dbin(p[i,k], n[i,k])   # Binomial likelihood
      logit(p[i,k]) <- mu[i] + delta[i,k]  # Model for linear predictor
      rhat[i,k] <- p[i,k] * n[i,k]   # Expected value of the numerators
      dev[i,k] <- 2 * (r[i,k] * (log(r[i,k]) - log(rhat[i,k]))   # Deviance contribution
        + (n[i,k] - r[i,k]) * (log(n[i,k] - r[i,k]) - log(n[i,k] - rhat[i,k])))
    }
    resdev[i] <- sum(dev[i,1:na[i]])   # Summed residual deviance contribution for this trial

    for (k in 2:na[i]) {   # Loop through arms
      delta[i,k] ~ dnorm(md[i,k],taud[i,k]) # trial-specific LOR distributions
```

```
      md[i,k] <- d[t[i,k]] - d[t[i,1]] + sw[i,k] # mean of treat effects distributions (with multi-arm trial correction)
      taud[i,k] <- tau_sd *2*(k-1)/k # precision of treat effects distributions (with multi-arm trial correction)
      w[i,k] <- (delta[i,k] - d[t[i,k]] + d[t[i,1]]) # adjustment for multi-arm RCTs
      sw[i,k] <- sum(w[i,1:(k-1)])/(k-1) # cumulative adjustment for multi-arm trials

    }
  }
  totresdev <- sum(resdev[])  # Total Residual Deviance
  d[1] <- 0  # Treatment effect zero for reference treatment

  # Dirichlet Process Stick-breaking Priors
  for (k in 2:nt) {
    # Assign treatment effects to cluster
    cluster[k] ~ dcat(p.dp[1:H])  # Cluster assignment
    d[k] <- theta[cluster[k]]  # Cluster-specific treatment effect
  }

  # Hyperparameters nu and s^2 must be provided as data.
  # For example:
  # nu <- 4
  # s2 <- 1
  # These define the inverse-gamma prior for c^2.

  # Draw c^2 from an Inv-Gamma(nu/2, nu*s^2/2)
  # JAGS doesn't have inverse-gamma directly, but:
  # If c^2 ~ IG(a,b) then (1/c^2) ~ Gamma(a, 1/b).
  # Here, a = nu/2 and b = (nu*s^2)/2, so the gamma rate = 2/(nu*s^2).
  u ~ dgamma(nu/2, 2/(nu*s2))
  c2 <- 1/u

  # Global scale tau ~ C+(0,1)
  # Half-Cauchy can be represented using a half-t with 1 d.o.f.
  tau_raw ~ dt(0,1,1)T(0,)
  tau <- tau_raw

  for (h in 1:H) {
    # Local scale lambda_j ~ C+(0,1)
    lambda[h] ~ dt(0,1,1)T(0,)
    lambda2[h] <- lambda[h]^2

    # Compute tilde lambda^2
    # tilde(lambda_j)^2 = (c^2 * lambda_j^2) / (c^2 + tau^2 * lambda_j^2)
    lam_tilde2[h] <- (c2 * lambda2[h]) / (c2 + tau^2 * lambda2[h])
```

```
    # beta_j ~ Normal(0, tau^2 * lam_tilde2[h])
    # variance = tau^2 * lam_tilde2[h]
    # precision = 1/(tau^2 * lam_tilde2[h])
    theta[h] ~ dnorm(0, 1/(tau^2 * lam_tilde2[h]))
  }

q.dp[1] ~ dbeta(1, dp.alp)
p.dp[1] <- q.dp[1]
r.dp[1] <- 1 - q.dp[1]
for (j in 2:(H-1)){
   q.dp[j] ~ dbeta(1, dp.alp)
   p.dp[j] <-       q.dp[j] * r.dp[j-1]
   r.dp[j] <- (1 - q.dp[j]) * r.dp[j-1]
}
p.dp[H] <- r.dp[H-1]
  #dp.alp ~ dgamma(2,1)
  dp.alp <- 1

 for (c in 1:(nt-1)) {
    for (k in (c+1):nt) {
      diff[c,k] <- (d[k]-d[c])
    }
  }
    sd ~ dunif(0, 2)
  tau_sd <- pow(sd, -2)
}
```

### 9.1.2.4. DP NMA with meta-regression

```
model {
  for(i in 1:ns) {   # Loop through studies
    w[i,1] <- 0   # Adjustment for multi-arm trials is zero for control arm
    delta[i,1] <- 0   # Treatment effect is zero for control arm
    mu[i] ~ dnorm(0, .001)   # Vague priors for all trial baselines

    for (k in 1:na[i]) {   # Loop through arms
      r[i,k] ~ dbin(p[i,k], n[i,k])   # Binomial likelihood
      logit(p[i,k]) <- mu[i] + delta[i,k] + (beta[t[i,k]]-beta[t[i,1]]) * (mu[i] - mx) # Model for linear predictor
      rhat[i,k] <- p[i,k] * n[i,k]   # Expected value of the numerators
      dev[i,k] <- 2 * (r[i,k] * (log(r[i,k]) - log(rhat[i,k]))   # Deviance contribution
        + (n[i,k] - r[i,k]) * (log(n[i,k] - r[i,k]) - log(n[i,k] - rhat[i,k])))
    }
    resdev[i] <- sum(dev[i,1:na[i]])   # Summed residual deviance contribution for this trial

    for (k in 2:na[i]) {   # Loop through arms
```

```
      delta[i,k] ~ dnorm(md[i,k],taud[i,k]) # trial-specific LOR distributions
      md[i,k] <- d[t[i,k]] - d[t[i,1]] + sw[i,k] # mean of treat effects distributions (with multi-arm trial correction)
      taud[i,k] <- tau_sd *2*(k-1)/k # precision of treat effects distributions (with multi-arm trial correction)
      w[i,k] <- (delta[i,k] - d[t[i,k]] + d[t[i,1]]) # adjustment for multi-arm RCTs
      sw[i,k] <- sum(w[i,1:(k-1)])/(k-1) # cumulative adjustment for multi-arm trials

    }
  }
  totresdev <- sum(resdev[])  # Total Residual Deviance
  d[1] <- 0  # Treatment effect zero for reference treatment

  # Dirichlet Process Stick-breaking Priors
  for (k in 2:nt) {
    # Assign treatment effects to cluster
    cluster[k] ~ dcat(p.dp[1:H])  # Cluster assignment
    d[k] <- theta[cluster[k]]  # Cluster-specific treatment effect
    beta[k] <- beta_theta[cluster[k]]
  }

  beta[1] <- 0

  # Hyperparameters nu and s^2 must be provided as data.
  # For example:
  # nu <- 4
  # s2 <- 1
  # These define the inverse-gamma prior for c^2.

  # Draw c^2 from an Inv-Gamma(nu/2, nu*s^2/2)
  # JAGS doesn't have inverse-gamma directly, but:
  # If c^2 ~ IG(a,b) then (1/c^2) ~ Gamma(a, 1/b).
  # Here, a = nu/2 and b = (nu*s^2)/2, so the gamma rate = 2/(nu*s^2).
  u ~ dgamma(nu/2, 2/(nu*s2))
  c2 <- 1/u

  # Global scale tau ~ C+(0,1)
  # Half-Cauchy can be represented using a half-t with 1 d.o.f.
  tau_raw ~ dt(0,1,1)T(0,)
  tau <- tau_raw

  for (h in 1:H) {
    # Local scale lambda_j ~ C+(0,1)
    lambda[h] ~ dt(0,1,1)T(0,)
    lambda2[h] <- lambda[h]^2
```

```
      # Compute tilde lambda^2
      # tilde(lambda_j)^2 = (c^2 * lambda_j^2) / (c^2 + tau^2 * lambda_j^2)
      lam_tilde2[h] <- (c2 * lambda2[h]) / (c2 + tau^2 * lambda2[h])

      # beta_j ~ Normal(0, tau^2 * lam_tilde2[h])
      # variance = tau^2 * lam_tilde2[h]
      # precision = 1/(tau^2 * lam_tilde2[h])
      theta[h] ~ dnorm(0, 1/(tau^2 * lam_tilde2[h]))
      beta_theta[h] <- B
   }

   B ~ dnorm(0, 0.001)

q.dp[1] ~ dbeta(1, dp.alp)
p.dp[1] <- q.dp[1]
r.dp[1] <- 1 - q.dp[1]
for (j in 2:(H-1)){
   q.dp[j] ~ dbeta(1, dp.alp)
   p.dp[j] <-        q.dp[j] * r.dp[j-1]
   r.dp[j] <- (1 - q.dp[j]) * r.dp[j-1]
}
p.dp[H] <- r.dp[H-1]
  #dp.alp ~ dunif(0, 10)
  #dp.alp ~ dgamma(2,1)
 dp.alp <- 1

 for (c in 1:(nt-1)) {
    for (k in (c+1):nt) {
       diff[c,k] <- (d[k]-d[c])
    }
  }
    sd ~ dunif(0, 2)
  tau_sd <- pow(sd, -2)

}
```

## 9.2. Pain case study

### 9.2.1. Data

```
list(t = structure(c(1L, 1L, 1L, 1L, 1L, 1L, 1L, 1L, 1L, 1L,
1L, 1L, 1L, 1L, 1L, 1L, 1L, 1L, 1L, 1L, 1L, 1L, 1L, 1L, 1L, 1L,
1L, 1L, 1L, 1L, 1L, 1L, 1L, 1L, 1L, 1L, 1L, 1L, 1L, 1L, 1L, 1L,
1L, 1L, 1L, 1L, 2L, 1L, 1L, 1L, 2L, 1L, 1L, 1L, 1L, 1L, 3L, 3L,
2L, 3L, 3L, 4L, 3L, 3L, 2L, 2L, 3L, 2L, 3L, 3L, 2L, 3L, 4L, 3L,
4L, 4L, 4L, 3L, 3L, 2L, 3L, 3L, 3L, 3L, 3L, 3L, 2L, 3L, 2L, 4L,
3L, 3L, 3L, 3L, 3L, 4L, 4L, 2L, 4L, 3L, 4L, 4L, 3L, 3L, 3L, 3L,
3L, 4L, 4L, 2L, 3L, 4L, NA, NA, 3L, NA, NA, NA, NA, NA, 3L, NA,
NA, NA, NA, NA, NA, 4L, NA, NA, NA, NA, NA, NA, NA, NA, NA, NA,
NA, NA, NA, NA, 3L, NA, NA, NA, NA, NA, NA, NA, NA, NA, NA, NA,
```

```
NA, NA, NA, NA, NA, NA, NA, NA, NA, NA, NA, NA, NA, NA), dim = c(56L,
3L), dimnames = list(NULL, c("t_1", "t_2", "t_3"))), na = c(2L,
2L, 3L, 2L, 2L, 2L, 2L, 2L, 3L, 2L, 2L, 2L, 2L, 2L, 2L, 3L, 2L,
2L, 2L, 2L, 2L, 2L, 2L, 2L, 2L, 2L, 2L, 2L, 2L, 3L, 2L, 2L,
2L, 2L, 2L, 2L, 2L, 2L, 2L, 2L, 2L, 2L, 2L, 2L, 2L, 2L, 2L,
2L, 2L, 2L, 2L, 2L, 2L, 2L), ns = 56L, nt = 4L, H = 4L, nu = 1,
    s2 = 1, y = structure(c(58.1, 44.8, 32.9, 51.6, 14.9, 72,
    55, 93, 54.9, 43.1, 64.2, 43.3, 59, 30.8, 21.1, 55.1, 17.4,
    36.5, 34.9, 43.5, 57.5, 30.9, 48.2, 20.1, 80.4, 38, 44, 21.7,
    51.6, 58.75, 66.7, 78, 59.5, 51, 38.2, 39.2, 51, 29.7, 22.1,
    36.6, 35.5, 12.45, 27, 8.55, 45.2, 47, 54.5, 56.5, 31.6,
    37, 65, 31.3, 44.2, 57.4, 19.5, 141.5, 44.01, 44.6, 28, 41.34,
    17.4, 54, 43, 63, 35, 34.5, 39.6, 26, 38, 20.9, 16.8, 36.6,
    12.6, 23.53, 25.6, 33.97, 40.35, 17.4, 39.1, 12.1, 32.4,
    32, 31.96, 15, 36.3, 30, 61.1, 53.97, 50.3, 33.51, 33.2,
    34.6, 33, 23, 8, 26.8, 25.5, 11.25, 12.5, 5.4, 28, 25.54,
    44.1, 46.05, 28.33, 27, 58, 25.93, 35.1, 39.56, 16.9, 73.56,
    NA, NA, 25.7, NA, NA, NA, NA, NA, 32.7, NA, NA, NA, NA, NA,
    NA, 35.2, NA, NA, NA, NA, NA, NA, NA, NA, NA, NA, NA, NA,
    NA, NA, 36, NA, NA, NA, NA, NA, NA, NA, NA, NA, NA, NA, NA,
    NA, NA, NA, NA, NA, NA, NA, NA, NA, NA, NA), dim = c(56L,
    3L), dimnames = list(NULL, c("y_1", "y_2", "y_3"))), se = structure(c(3.0
649781,
    6.1967734, 6.506612, 3.9244426, 4.7750393, 6.6018197, 4.0852974,
    1.3416408, 6.1755663, 2.9029296, 4.3705894, 3.3387337, 5.0002116,
    2.8001488, 2.008316, 3.0983867, 1.634119, 5.2414375, 2.3143304,
    2.3559785, 3.9480256, 5.1847956, 4.5080931, 1.8670719, 11.198086,
    2.8571429, 3.8758512, 2.899796, 7.5100409, 17.000079, 4.472136,
    8.9991123, 8.2957125, 6.363961, 4.902607, 1.725, 5.4738594,
    0.8497058, 5.1961524, 2.2979701, 2.3004347, 1.5697197, 1.6099689,
    1.1582832, 2.7110883, 5.891883, 5.8175381, 4.7992197, 0.8336148,
    3.354102, 6.3960215, 2.4373141, 0.82, 7.2527051, 2.1430508,
    17.654099, 2.1036088, 5.344717, 5.2414375, 3.3120136, 4.0160926,
    5.4738594, 3.1037612, 1.3416408, 4.1641326, 2.3186922, 3.0039847,
    2.6622583, 3.9998101, 2.2211609, 1.5336232, 2.32379, 1.1867322,
    2.3583064, 1.3237522, 1.5830774, 3.307234, 3.653385, 3.1753903,
    1.5276032, 6.4988661, 2.4271195, 2.3709019, 1.9052559, 6.6064793,
    9.5082573, 5.1429563, 4.8887125, 8.02, 3.3225035, 3.9833682,
    1.6124636, 3.584417, 0.5391683, 2.0801257, 1.7870153, 1.5153657,
    1.8827692, 0.910736, 0.9615092, 1.8203022, 1.9240607, 4.88,
    4.1389736, 0.6392041, 2.9104275, 5.2128604, 1.3381157, 0.7,
    3.2734082, 1.6782928, 9.6391968, NA, NA, 4.3893811, NA, NA,
    NA, NA, NA, 6.1268263, NA, NA, NA, NA, NA, NA, 2.1430508,
    NA, NA, NA, NA, NA, NA, NA, NA, NA, NA, NA, NA, NA, 4.0249224,
    NA, NA, NA, NA, NA, NA, NA, NA, NA, NA, NA, NA, NA, NA,
    NA, NA, NA, NA, NA, NA, NA, NA, NA, NA), dim = c(56L, 3L), dimnames = lis
t(
        NULL, c("se_1", "se_2", "se_3"))), mx = 45.2583333333333)
```

## 9.2.2. Model code

### 9.2.2.1. Standard NMA

```
model {
  for(i in 1:ns) {  # Loop through studies
    delta[i,1] <- 0  # Treatment effect is zero for control arm
    mu[i] ~ dnorm(0, .001)  # Vague priors for all trial baselines
    w[i,1] <- 0
    for (k in 1:na[i]) {  # Loop through arms
      prec[i,k] <- pow(se[i,k], -2)
      y[i,k] ~ dnorm(theta_y[i,k],prec[i,k])
      theta_y[i,k] <- mu[i] + delta[i,k]
   dev[i,k] <- (y[i,k]-theta_y[i,k])*(y[i,k]-theta_y[i,k])*prec[i,k] #Deviance contribution
    }
    resdev[i] <- sum(dev[i,1:na[i]])  # Summed residual deviance contribution for this trial

   for (k in 2:na[i]) {  # Loop through arms
      delta[i,k] ~ dnorm(md[i,k],taud[i,k]) # trial-specific LOR distributions
      md[i,k] <- d[t[i,k]] - d[t[i,1]] + sw[i,k] # mean of treat effects distributions (with multi-arm trial correction)
      taud[i,k] <- tau_sd *2*(k-1)/k # precision of treat effects distributions (with multi-arm trial correction)
      w[i,k] <- (delta[i,k] - d[t[i,k]] + d[t[i,1]]) # adjustment for multi-arm RCTs
      sw[i,k] <- sum(w[i,1:(k-1)])/(k-1) # cumulative adjustment for multi-arm trials

    }
  }
  totresdev <- sum(resdev[])  # Total Residual Deviance
  d[1] <- 0  # Treatment effect zero for reference treatment

  # Dirichlet Process Stick-breaking Priors
  for (k in 2:nt) {
    # Local scale lambda_j ~ C+(0,1)
    lambda[k] ~ dt(0,1,1)T(0,)
    lambda2[k] <- lambda[k]^2

    # Compute tilde lambda^2
    # tilde(lambda_j)^2 = (c^2 * lambda_j^2) / (c^2 + tau^2 * lambda_j^2)
    lam_tilde2[k] <- (c2 * lambda2[k]) / (c2 + tau^2 * lambda2[k])

    # beta_j ~ Normal(0, tau^2 * lam_tilde2[k])
    # variance = tau^2 * lam_tilde2[k]
    # precision = 1/(tau^2 * lam_tilde2[k])
    d[k] ~ dnorm(0, 1/(tau^2 * lam_tilde2[k]))
```

```
  }

  # Hyperparameters nu and s^2 must be provided as data.
  # For example:
  # nu <- 4
  # s2 <- 1
  # These define the inverse-gamma prior for c^2.

  # Draw c^2 from an Inv-Gamma(nu/2, nu*s^2/2)
  # JAGS doesn't have inverse-gamma directly, but:
  # If c^2 ~ IG(a,b) then (1/c^2) ~ Gamma(a, 1/b).
  # Here, a = nu/2 and b = (nu*s^2)/2, so the gamma rate = 2/(nu*s^2).
  u ~ dgamma(nu/2, 2/(nu*s2))
  c2 <- 1/u

  # Global scale tau ~ C+(0,1)
  # Half-Cauchy can be represented using a half-t with 1 d.o.f.
  tau_raw ~ dt(0,1,1)T(0,)
  tau <- tau_raw

 for (c in 1:(nt-1)) {
   for (k in (c+1):nt) {
     diff[c,k] <- (d[k]-d[c])
   }
 }
  sd ~ dunif(0, 10)
  tau_sd <- pow(sd, -2)
}
```

### 9.2.2.2. Standard NMA with meta-regression

```
model {
  for(i in 1:ns) {   # Loop through studies
    delta[i,1] <- 0  # Treatment effect is zero for control arm
    mu[i] ~ dnorm(0, .001)  # Vague priors for all trial baselines
    w[i,1] <- 0
    for (k in 1:na[i]) {   # Loop through arms
      prec[i,k] <- pow(se[i,k], -2)
      y[i,k] ~ dnorm(theta_y[i,k],prec[i,k])
      theta_y[i,k] <- mu[i] + delta[i,k] + (beta[t[i,k]]-beta[t[i,1]]) * (mu[i] - mx) # Model for linear predictor
   dev[i,k] <- (y[i,k]-theta_y[i,k])*(y[i,k]-theta_y[i,k])*prec[i,k] #Deviance contribution
    }
    resdev[i] <- sum(dev[i,1:na[i]])   # Summed residual deviance contribution for this trial
```

```
   for (k in 2:na[i]) {  # Loop through arms
      delta[i,k] ~ dnorm(md[i,k],taud[i,k]) # trial-specific LOR distributions
      md[i,k] <- d[t[i,k]] - d[t[i,1]] + sw[i,k] # mean of treat effects distributions (with multi-arm trial correction)
      taud[i,k] <- tau_sd *2*(k-1)/k # precision of treat effects distributions (with multi-arm trial correction)
      w[i,k] <- (delta[i,k] - d[t[i,k]] + d[t[i,1]]) # adjustment for multi-arm RCTs
      sw[i,k] <- sum(w[i,1:(k-1)])/(k-1) # cumulative adjustment for multi-arm trials

    }
  }
  totresdev <- sum(resdev[])  # Total Residual Deviance
  d[1] <- 0  # Treatment effect zero for reference treatment
  beta[1] <- 0
  for (k in 2:nt) {
    d[k] ~ dnorm(0, 0.001)
    beta[k] <- B
  }

  # vague priors for treatment effects
  B ~ dnorm(0,.001)
  sd ~ dunif(0, 10)
  tau_sd <- pow(sd, -2)

   for (c in 1:(nt-1)) {
    for (k in (c+1):nt) {
      diff[c,k] <- (d[k]-d[c])
    }
   }

}
```

### 9.2.2.3. DP NMA

```
model {
   for(i in 1:ns) {  # Loop through studies
     delta[i,1] <- 0  # Treatment effect is zero for control arm
     mu[i] ~ dnorm(0, .001)  # Vague priors for all trial baselines
     w[i,1] <- 0
     for (k in 1:na[i]) {  # Loop through arms
       prec[i,k] <- pow(se[i,k], -2)
       y[i,k] ~ dnorm(theta_y[i,k],prec[i,k])
       theta_y[i,k] <- mu[i] + delta[i,k]
   dev[i,k] <- (y[i,k]-theta_y[i,k])*(y[i,k]-theta_y[i,k])*prec[i,k] #Deviance contribution
     }
     resdev[i] <- sum(dev[i,1:na[i]])  # Summed residual deviance contribution for this trial
```

```
   for (k in 2:na[i]) {  # Loop through arms
      delta[i,k] ~ dnorm(md[i,k],taud[i,k]) # trial-specific LOR distributions
      md[i,k] <- d[t[i,k]] - d[t[i,1]] + sw[i,k] # mean of treat effects distributions (with multi-arm trial correction)
      taud[i,k] <- tau_sd *2*(k-1)/k # precision of treat effects distributions (with multi-arm trial correction)
      w[i,k] <- (delta[i,k] - d[t[i,k]] + d[t[i,1]]) # adjustment for multi-arm RCTs
      sw[i,k] <- sum(w[i,1:(k-1)])/(k-1) # cumulative adjustment for multi-arm trials

    }
  }
  totresdev <- sum(resdev[])  # Total Residual Deviance
  d[1] <- 0  # Treatment effect zero for reference treatment

  # Dirichlet Process Stick-breaking Priors
  for (k in 2:nt) {
    # Assign treatment effects to cluster
    cluster[k] ~ dcat(p.dp[1:H])  # Cluster assignment
    d[k] <- theta[cluster[k]]  # Cluster-specific treatment effect
  }

  # Hyperparameters nu and s^2 must be provided as data.
  # For example:
  # nu <- 4
  # s2 <- 1
  # These define the inverse-gamma prior for c^2.

  # Draw c^2 from an Inv-Gamma(nu/2, nu*s^2/2)
  # JAGS doesn't have inverse-gamma directly, but:
  # If c^2 ~ IG(a,b) then (1/c^2) ~ Gamma(a, 1/b).
  # Here, a = nu/2 and b = (nu*s^2)/2, so the gamma rate = 2/(nu*s^2).
  u ~ dgamma(nu/2, 2/(nu*s2))
  c2 <- 1/u

  # Global scale tau ~ C+(0,1)
  # Half-Cauchy can be represented using a half-t with 1 d.o.f.
  tau_raw ~ dt(0,1,1)T(0,)
  tau <- tau_raw

  for (h in 1:H) {
    # Local scale lambda_j ~ C+(0,1)
    lambda[h] ~ dt(0,1,1)T(0,)
    lambda2[h] <- lambda[h]^2
```

```
    # Compute tilde lambda^2
    # tilde(lambda_j)^2 = (c^2 * lambda_j^2) / (c^2 + tau^2 * lambda_j^2)
    lam_tilde2[h] <- (c2 * lambda2[h]) / (c2 + tau^2 * lambda2[h])

    # beta_j ~ Normal(0, tau^2 * lam_tilde2[h])
    # variance = tau^2 * lam_tilde2[h]
    # precision = 1/(tau^2 * lam_tilde2[h])
    theta[h] ~ dnorm(0, 1/(tau^2 * lam_tilde2[h]))
  }

q.dp[1] ~ dbeta(1, dp.alp)
p.dp[1] <- q.dp[1]
r.dp[1] <- 1 - q.dp[1]
for (j in 2:(H-1)){
   q.dp[j] ~ dbeta(1, dp.alp)
   p.dp[j] <-        q.dp[j] * r.dp[j-1]
   r.dp[j] <- (1 - q.dp[j]) * r.dp[j-1]
}
p.dp[H] <- r.dp[H-1]
  #dp.alp ~ dgamma(2,1)
  dp.alp <- 1

 for (c in 1:(nt-1)) {
   for (k in (c+1):nt) {
     diff[c,k] <- (d[k]-d[c])
   }
  }
    sd ~ dunif(0, 10)
  tau_sd <- pow(sd, -2)
}
```

### 9.2.2.4. DP NMA with meta-regression

```
model {
  for(i in 1:ns) {   # Loop through studies
    delta[i,1] <- 0  # Treatment effect is zero for control arm
    mu[i] ~ dnorm(0, .001)  # Vague priors for all trial baselines
    w[i,1] <- 0
    for (k in 1:na[i]) {   # Loop through arms
      prec[i,k] <- pow(se[i,k], -2)
      y[i,k] ~ dnorm(theta_y[i,k],prec[i,k])
      theta_y[i,k] <- mu[i] + delta[i,k] + (beta[t[i,k]]-beta[t[i,1]]) * (mu[i] - mx) # Model for linear predictor
   dev[i,k] <- (y[i,k]-theta_y[i,k])*(y[i,k]-theta_y[i,k])*prec[i,k] #Deviance contribution
    }
    resdev[i] <- sum(dev[i,1:na[i]])   # Summed residual deviance contribution for this trial
```

```
   for (k in 2:na[i]) {  # Loop through arms
      delta[i,k] ~ dnorm(md[i,k],taud[i,k]) # trial-specific LOR distributions
      md[i,k] <- d[t[i,k]] - d[t[i,1]] + sw[i,k] # mean of treat effects distributions (with multi-arm trial correction)
      taud[i,k] <- tau_sd *2*(k-1)/k # precision of treat effects distributions (with multi-arm trial correction)
      w[i,k] <- (delta[i,k] - d[t[i,k]] + d[t[i,1]]) # adjustment for multi-arm RCTs
      sw[i,k] <- sum(w[i,1:(k-1)])/(k-1) # cumulative adjustment for multi-arm trials

    }
  }
  totresdev <- sum(resdev[])  # Total Residual Deviance
  d[1] <- 0  # Treatment effect zero for reference treatment

  # Dirichlet Process Stick-breaking Priors
  for (k in 2:nt) {
    # Assign treatment effects to cluster
    cluster[k] ~ dcat(p.dp[1:H])  # Cluster assignment
    d[k] <- theta[cluster[k]]  # Cluster-specific treatment effect
    beta[k] <- beta_theta[cluster[k]]
  }

  beta[1] <- 0

  # Hyperparameters nu and s^2 must be provided as data.
  # For example:
  # nu <- 4
  # s2 <- 1
  # These define the inverse-gamma prior for c^2.

  # Draw c^2 from an Inv-Gamma(nu/2, nu*s^2/2)
  # JAGS doesn't have inverse-gamma directly, but:
  # If c^2 ~ IG(a,b) then (1/c^2) ~ Gamma(a, 1/b).
  # Here, a = nu/2 and b = (nu*s^2)/2, so the gamma rate = 2/(nu*s^2).
  u ~ dgamma(nu/2, 2/(nu*s2))
  c2 <- 1/u

  # Global scale tau ~ C+(0,1)
  # Half-Cauchy can be represented using a half-t with 1 d.o.f.
  tau_raw ~ dt(0,1,1)T(0,)
  tau <- tau_raw

  for (h in 1:H) {
    # Local scale lambda_j ~ C+(0,1)
    lambda[h] ~ dt(0,1,1)T(0,)
    lambda2[h] <- lambda[h]^2
```

```
      # Compute tilde lambda^2
      # tilde(lambda_j)^2 = (c^2 * lambda_j^2) / (c^2 + tau^2 * lambda_j^2)
      lam_tilde2[h] <- (c2 * lambda2[h]) / (c2 + tau^2 * lambda2[h])
      
      # beta_j ~ Normal(0, tau^2 * lam_tilde2[h])
      # variance = tau^2 * lam_tilde2[h]
      # precision = 1/(tau^2 * lam_tilde2[h])
      theta[h] ~ dnorm(0, 1/(tau^2 * lam_tilde2[h]))
      beta_theta[h] <- B
    }
    
  B ~ dnorm(0, 0.001)
  
q.dp[1] ~ dbeta(1, dp.alp)
p.dp[1] <- q.dp[1]
r.dp[1] <- 1 - q.dp[1]
for (j in 2:(H-1)){
   q.dp[j] ~ dbeta(1, dp.alp)
   p.dp[j] <-        q.dp[j] * r.dp[j-1]
   r.dp[j] <- (1 - q.dp[j]) * r.dp[j-1]
}
p.dp[H] <- r.dp[H-1]
 dp.alp <- 1
 
 for (c in 1:(nt-1)) {
    for (k in (c+1):nt) {
      diff[c,k] <- (d[k]-d[c])
    }
  }
    sd ~ dunif(0, 10)
   tau_sd <- pow(sd, -2)
   
}
```

36